\documentclass[aps,prc,reprint,superscriptaddress,nofootinbib,floatfix]{revtex4-2}

\usepackage{amsmath,amssymb,bm}
\usepackage{graphicx}
\graphicspath{{figures_eps/}}
\DeclareGraphicsExtensions{.eps}
\usepackage{booktabs}
\usepackage{multirow}
\usepackage[table]{xcolor}

\newcommand{\hOm}{\hbar\Omega}

\begin{document}

\title{Finite-spectrum Lorentz integral transform calculation of the $^{4}$He photoabsorption cross section in the no-core shell model}

\author{P. Yin}
\email{pengyin@iastate.edu}
\affiliation{College of Physics and Engineering, Henan University of Science and Technology, Luoyang 471023, China}
\affiliation{State Key Laboratory of Heavy Ion Science and Technology, Institute of Modern Physics, Chinese Academy of Sciences, Lanzhou 730000, China}

\author{H. T. Zhao}
\affiliation{College of Physics and Engineering, Henan University of Science and Technology, Luoyang 471023, China}

\author{C. Y. Zhai}
\affiliation{College of Physics and Engineering, Henan University of Science and Technology, Luoyang 471023, China}

\author{J. P. Vary}
\affiliation{Department of Physics and Astronomy, Iowa State University, Ames, IA 50011, USA}

\author{H. Li}
\email{lihe2007@impcas.ac.cn}
\affiliation{CAS Key Laboratory of High Precision Nuclear Spectroscopy, Institute of Modern Physics, Chinese Academy of Sciences, Lanzhou 730000, China}
\affiliation{School of Nuclear Science and Technology, University of Chinese Academy of Sciences, Beijing 100049, China}

\author{J. M. Dong}
\affiliation{CAS Key Laboratory of High Precision Nuclear Spectroscopy, Institute of Modern Physics, Chinese Academy of Sciences, Lanzhou 730000, China}
\affiliation{School of Nuclear Science and Technology, University of Chinese Academy of Sciences, Beijing 100049, China}

\author{H. J. Ong}
\affiliation{CAS Key Laboratory of High Precision Nuclear Spectroscopy, Institute of Modern Physics, Chinese Academy of Sciences, Lanzhou 730000, China}
\affiliation{School of Nuclear Science and Technology, University of Chinese Academy of Sciences, Beijing 100049, China}

\author{X. Zhao}
\affiliation{CAS Key Laboratory of High Precision Nuclear Spectroscopy, Institute of Modern Physics, Chinese Academy of Sciences, Lanzhou 730000, China}
\affiliation{School of Nuclear Science and Technology, University of Chinese Academy of Sciences, Beijing 100049, China}

\author{P. J. Fasano}
\affiliation{Department of Physics and Astronomy, University of Notre Dame, Notre Dame, IN 46556-5670, USA}

\author{A. M. Shirokov}
\affiliation{Skobeltsyn Institute of Nuclear Physics, Lomonosov Moscow State University, Moscow 119991, Russia}

\author{J. Chen}
\affiliation{College of Science, Southern University of Science and Technology, Shenzhen 518055, Guangdong, China}

\author{D. Y. Tao}
\affiliation{Key Laboratory of Nuclear Physics and Ion-Beam Application (MoE), Institute of Modern Physics, Fudan University,  Shanghai 200433, China}
\affiliation{Shanghai Research Center for Theoretical Nuclear Physics, NSFC and Fudan University, Shanghai 200438, China}

\author{B. Zhou}
\affiliation{Key Laboratory of Nuclear Physics and Ion-Beam Application (MoE), Institute of Modern Physics, Fudan University,  Shanghai 200433, China}
\affiliation{Shanghai Research Center for Theoretical Nuclear Physics, NSFC and Fudan University, Shanghai 200438, China}

\author{C. Ji}
\affiliation{Key Laboratory of Quark and Lepton Physics, Institute of Particle Physics,
Central China Normal University, Wuhan 430079, China}
\affiliation{Southern Center for Nuclear-Science Theory (SCNT), Institute of Modern Physics,
Chinese Academy of Sciences, Huizhou 516000, Guangdong Province, China}

\begin{abstract}
We develop and validate a finite-spectrum implementation of the Lorentz integral transform (LIT) within the \textit{ab initio} no-core shell model (NCSM) for calculating the photoabsorption cross section of $^4$He.
A large set of $1^-$ eigenstates is explicitly calculated in the NCSM, and the LIT is constructed from their excitation energies and the corresponding $E1$ transition strengths.
This finite-spectrum approach is complementary to conventional inhomogeneous-equation and Lanczos-based implementations of the LIT method  for photoabsorption cross sections.
Using the Daejeon16 interaction, we extract the photoabsorption cross section and examine its stability with respect to the model-space truncation, excitation-energy cutoff, and LIT parameters.
The reliability of the finite-spectrum extraction is assessed by comparing the $E1$ polarizability and bremsstrahlung sum rule obtained from the discrete NCSM spectrum with the same quantities obtained by integrating the extracted cross section.
The extracted cross section captures the principal features of the available $^4$He photonuclear data in the giant-dipole-resonance region and is consistent, in the low-energy rise and main-peak region, with earlier chiral-interaction NCSM--LIT results obtained from Lanczos-based evaluations, while the present calculation with the Daejeon16 interaction exhibits a more pronounced high-energy shoulder.
The present work provides a controlled finite-spectrum NCSM--LIT route from explicitly calculated many-body eigenstates and transition strengths to photoabsorption cross sections.
\end{abstract}

\maketitle

\section{Introduction}
\label{sec:introduction}

Photonuclear absorption provides a clean electromagnetic probe of nuclear dynamics, owing to the perturbative and well-characterized coupling of photons to the nuclear electromagnetic current~\cite{EfrosLITReview,LeidemannOrlandini2013,BaccaPastore2014,BermanFultz1975}.
The photoabsorption cross section in the giant-dipole region is governed predominantly by isovector $E1$ strength; its energy dependence and sum rules are sensitive to the nuclear Hamiltonian and to correlations induced by the Hamiltonian~\cite{BermanFultz1975,HarakehWoude2001,GazitSumRules2006}.
These observables therefore provide stringent benchmarks for \textit{ab initio} descriptions of electromagnetic response in light nuclei~\cite{EfrosLITReview,LeidemannOrlandini2013,BaccaPastore2014,Stetcu2007}.
Accurate calculations of photonuclear cross sections are important for testing microscopic Hamiltonians and many-body methods in the electromagnetic response of light nuclei~\cite{QuaglioniNavratil2007,BaccaPastore2014}.

The $^{4}$He nucleus is a particularly useful system for such studies.
Its small mass permits high-precision \textit{ab initio} calculations, while its electromagnetic observables remain sensitive to details of the nuclear Hamiltonian.
For instance, this sensitivity has recently been highlighted in the isoscalar monopole sector, where precision measurements of the $0_1^+\rightarrow 0_2^+$ transition form factor~\cite{KegelMonopole2023} exposed a low-energy challenge for microscopic descriptions of $^{4}$He dynamics~\cite{BaccaMonopole2013} and motivated renewed theoretical work~\cite{Michel2023,Meissner2024,Viviani2024,YinMonopole2025}.
In the isovector $E1$ sector, $^{4}$He photoabsorption has likewise served as a long-standing benchmark: previous \textit{ab initio} calculations have shown sensitivity of the cross section to the underlying nuclear interaction, including three-nucleon-force effects~\cite{Gazit2006,QuaglioniNavratil2007,BaccaPastore2014}, whereas the available measurements in the giant-dipole-resonance region still show appreciable spread~\cite{ArkatovData,Calarco1983,Shima2005,Nilsson2007,Raut2012,Tornow2012,Murata2023}.
The $^{4}$He photoabsorption cross section therefore provides a demanding test case for assessing whether finite-basis many-body spectral information can be converted into a stable and smooth energy-dependent photonuclear cross section.

The Lorentz integral transform (LIT) method was introduced to calculate nuclear response functions without the explicit construction of continuum wave functions~\cite{EfrosLIT,EfrosLITReview,LeidemannOrlandini2013}.
In the conventional inhomogeneous-equation formulation, the LIT is obtained from a localized, bound-state-like solution of a source equation~\cite{EfrosLITReview,LeidemannOrlandini2013}.
The inversion of the transform has been studied extensively, including regularization procedures and attainable energy resolution~\cite{Andreasi2005,BarneaInversion2010,Leidemann2015}.
Lanczos evaluations of the LIT provide a complementary route in which the transform is generated from the tridiagonal representation of the Hamiltonian in a Krylov basis and can be implemented efficiently in large many-body spaces~\cite{MarchisioLanczos,EfrosLITReview}.
LIT calculations have been carried out within various \textit{ab initio} approaches, such as hyperspherical harmonics method~\cite{Bacca2002,Gazit2006,Barnea2004,BaccaPastore2014}, no-core shell model (NCSM)~\cite{QuaglioniNavratil2007,Stetcu2007,Schuster2014}, and coupled-cluster theory~\cite{BaccaCC2013,Miorelli2016,OrlandiniCC2014}.

A recent direct \textit{ab initio} NCSM calculation showed that the inverse-energy-weighted $E1$ sum rule of $^{4}$He can be obtained with excellent convergence by explicitly computing a large number of discretized $1^-$ states~\cite{Yin2024}.
That result raises a natural question: whether the same large-scale NCSM spectral information can be used not only for an integrated sum rule, but also for an energy-dependent photonuclear observable.
In the present work we address this question by developing a finite-spectrum NCSM-LIT approach, in which the transform is constructed directly from the $1^-$ excitation energies and $E1$ transition strengths calculated with the NCSM and then inverted to obtain the $^{4}$He photoabsorption cross section.

This finite-spectrum implementation complements standard inhomogeneous-equation and Lanczos LIT approaches, in which the transform is evaluated without retaining the individual final-state spectrum.
It instead uses the resolved large-scale \textit{ab initio} NCSM spectrum as direct input to the LIT inversion, thereby connecting two levels of information obtained in a finite harmonic-oscillator (HO) basis: the discrete $E1$ spectrum and the continuum observable measured in photonuclear reactions.
A primary goal of this work is to demonstrate how this connection provides a controlled route from microscopic nuclear structure to an energy-dependent photonuclear cross section.

\section{Theoretical Method}
\label{sec:method}

\begin{figure*}[t!]
\centering
\includegraphics[width=0.96\textwidth]{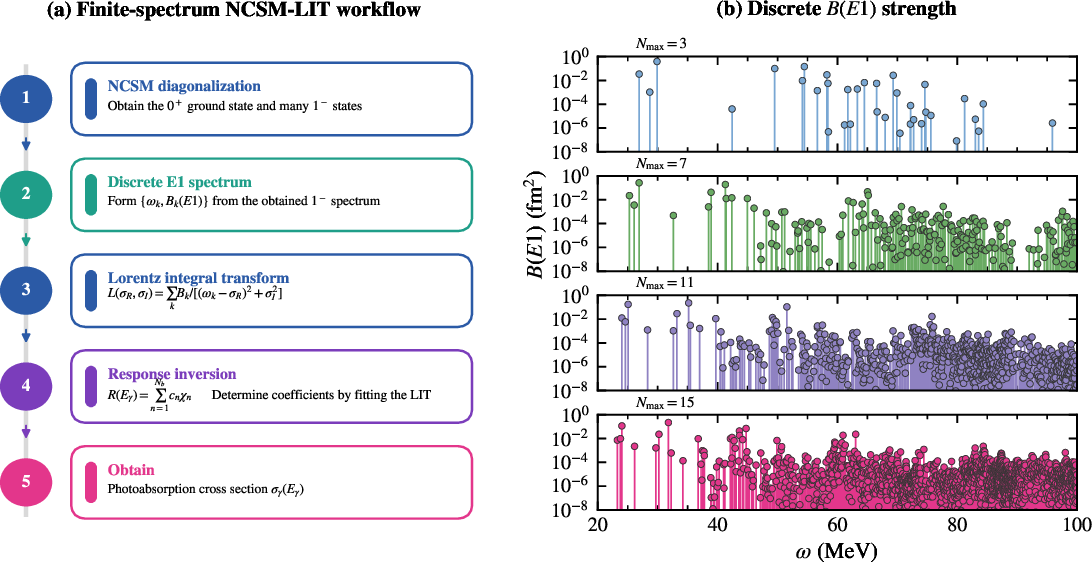}
\caption{
Method overview and finite NCSM $B(E1)$ spectra.
Panel (a) shows the finite-spectrum NCSM-LIT workflow.
Panel (b) shows representative NCSM $B(E1)$ strength distributions at $\hOm=15$ MeV for increasing $N_{\max}$.
Each vertical line is a calculated $0^+\!\rightarrow 1^-$ transition retained in the finite spectrum used to construct the transform.
}
\label{fig:workflow}
\end{figure*}

We implement the finite-spectrum NCSM-LIT strategy following the sequence shown in Fig.~\ref{fig:workflow}(a).
The calculation starts from an NCSM diagonalization, which provides the $0^+$ ground state and a large finite set of $J^\pi=1^-$ eigenstates of $^4$He in the HO basis.
These states enter the LIT construction through their excitation energies and $E1$ transition matrix elements.
The inverted transform yields the energy-dependent $E1$ strength distribution, from which the total photoabsorption cross section is constructed.

We use the \textit{ab initio} NCSM~\cite{Barrett2013} to generate the many-body spectral input for the finite-spectrum NCSM-LIT calculation.
The NCSM has been extensively used recently in studies of $s$- and $p$-shell nuclei (see, e.g., Refs.~\cite{Maris2016,Maris2021,Maris2022,Yin2024,YinMonopole2025,LiCPC2024,LiPRC2024,HuangHalo2025}).
In the NCSM, all nucleons are active, and the chosen nuclear Hamiltonian is diagonalized in a truncated HO Slater-determinant basis.
The HO energy is denoted by $\hOm$, and the basis size is specified by $N_{\max}$, the maximum number of HO excitation quanta above the lowest Pauli-allowed configuration.
We use the Daejeon16 nucleon-nucleon ($NN$) interaction~\cite{Daejeon16} and follow the same NCSM spectroscopic setup as in the direct evaluation of the $^{4}$He $E1$ polarizability~\cite{Yin2024}.
The many-body diagonalizations and $E1$ matrix-element evaluations are performed in the MFDn-based NCSM framework~\cite{MarisMFDn2010,AktulgaMFDn2014,ShaoMFDn2018,MarisMFDnGPU2022,CookMFDn2022,FasanoMFDnTransitions2025}.

The diagonalizations are carried out in an $M$-scheme basis with fixed angular-momentum projection $M=\sum_i m_i$, isospin projection $M_T=\sum_i m_{t_i}$ and parity.
The $^{4}$He ground state is the normal-parity $J^\pi=0^+$ state and is obtained with even $N_{\max}$ ($N_{\max}=2, 4, 6,\cdots$).
According to the $E1$ selection rule, the relevant excited states have $J^\pi=1^-$; they are of abnormal parity and are obtained with odd $N_{\max}$ ($N_{\max}=3, 5, 7,\cdots$).
For simplicity, the results below are labeled by the odd $N_{\max}$ value used for the excited $1^-$ states, while the $0^+$ ground state is calculated with $N_{\max}-1$.
As in Ref.~\cite{Yin2024}, the $1^-$ states are selected in the $M=1$ sector.
States with $J>1$ are shifted away by adding the angular-momentum filter $H_{J^2}=\lambda(\mathbf{J}^2-2)$ during the diagonalization.
Here $\mathbf J=\sum_i\mathbf j_i$ is the total angular momentum, and we use $\lambda=25$ MeV.
This term leaves the $J=1$ states unchanged and shifts, for example, the $2^-$ and $3^-$ states upward by 100 and 250 MeV, respectively.
Spurious center-of-mass excitations are removed with the Lawson prescription~\cite{GloecknerLawson,Whitehead1977}.
The finite-spectrum character of the calculation arises because only a finite set of $1^-$ eigenstates is computed below a chosen excitation energy cutoff $\omega_{\rm cut}$.
As a representative case, at $N_{\max}=11$, $\hOm=20$ MeV, with $M=1$, the unfiltered negative-parity spectrum contains $2346$ states below $\omega_{\rm cut}=100$ MeV, of which $370$ are $1^-$ states.
For each model space and cutoff $\omega_{\rm cut}$ considered below, the resulting finite set of $1^-$ eigenstates provides the spectral input for the LIT construction.

For the $k$th retained $1^-_k$ state with excitation energy $\omega_k$, the
reduced $E1$ strength is
\begin{align}
B_k(E1)
&\equiv B(E1;0^+\rightarrow 1^-_k)  \nonumber\\
&=
\sum_{M_k\mu}
\left|
\langle J_kM_k|\hat D_{1\mu}|J_0M_0\rangle
\right|^2
\nonumber\\
&=
\frac{
\left|
\langle J_k\Vert\hat D_{1}\Vert J_0\rangle
\right|^2}
{2J_0+1},
\label{eq:BE1_definition}
\end{align}
where $J_0$ ($J_k$) and $M_0$ ($M_k$) are the total angular momentum and its
projection of the ground (excited) state, respectively, with $J_0=M_0=0$ for
the $^4$He ground state.
Here $\langle\cdots|\cdots|\cdots\rangle$ denotes an ordinary matrix
element between angular-momentum eigenstates, while
$\langle\cdots\Vert\cdots\Vert\cdots\rangle$ denotes the corresponding
reduced matrix element.
In the long-wavelength approximation, we employ the one-body $E1$ operator
\begin{equation}
\hat D_{1\mu}
=
\sum_{i=1}^{A} q_i r_iY_{1\mu}(\hat{\mathbf r}_i),
\label{eq:e1_operator}
\end{equation}
where $q_i$ is the charge number of nucleon $i$, and $\mathbf r_i$ is the
single-particle coordinate of nucleon $i$, with $r_i=|\mathbf r_i|$ and
$\hat{\mathbf r}_i=\mathbf r_i/r_i$.

In Fig.~\ref{fig:workflow}(b) we display the finite NCSM
spectral input used to construct the LIT, taking $\hOm=15$ MeV as an example.
The vertical lines correspond to the calculated $0^+\!\rightarrow 1^-$
transitions from the NCSM diagonalization for several $N_{\max}$ values, with
their positions and heights giving $\omega_k$ and $B_k(E1)$, respectively.
For each model space, the collection of pairs $\{\omega_k,B_k(E1)\}$ forms the
finite spectral input to the LIT, from which the smooth $E1$ response is
obtained by inversion afterwards.

The inclusive $E1$ response entering the photoabsorption cross section is written formally as
\begin{equation}
R(E_\gamma)
=
\sum_f
B(E1;0^+\rightarrow f)\,
\delta(E_\gamma-\omega_f),
\label{eq:response_general}
\end{equation}
where $E_\gamma$ is the photon energy, $\omega_f$ is the excitation energy of the final state $f$, and the sum runs over the allowed final states.
In the finite NCSM basis used here, this spectral representation is evaluated with the calculated $1^-$ eigenstates retained below the excitation-energy cutoff $\omega_{\rm cut}$,
\begin{equation}
R(E_\gamma)
\approx
\sum_{\omega_k\leq \omega_{\rm cut}}
B_k(E1)\,
\delta(E_\gamma-\omega_k).
\label{eq:discrete_response}
\end{equation}

The LIT generated from the finite NCSM input is then
\begin{align}
L(\sigma_R,\sigma_I)
&=
\int dE_\gamma\,
\frac{R(E_\gamma)}
{(E_\gamma-\sigma_R)^2+\sigma_I^2}
\nonumber\\
&\approx
\sum_{\omega_k\leq \omega_{\rm cut}}
\frac{B_k(E1)}
{(\omega_k-\sigma_R)^2+\sigma_I^2},
\label{eq:lit_discrete}
\end{align}
where $\sigma_R$ and $\sigma_I>0$ are the real and imaginary Lorentz parameters, respectively.

The inversion seeks a smooth response whose LIT reproduces Eq.~\eqref{eq:lit_discrete}.
We use the inversion basis form employed in the LIT method of Efros et al.~\cite{EfrosLIT,EfrosLITReview}, leading to the expansion
\begin{equation}
R(E_\gamma)
\approx
\sum_{n=1}^{N_b}
c_n\chi_n(E_\gamma-\omega_{\rm th}),
\label{eq:response_expansion}
\end{equation}
with
\begin{equation}
\chi_n(E)=
\begin{cases}
E^{3/2}\exp[-E/(n\beta)], & E\geq 0,\\
0, & E<0.
\end{cases}
\label{eq:basis_function}
\end{equation}
Here $\omega_{\rm th}$ is the calculated proton-emission threshold for the
$p+{}^3$H channel, obtained from the same NCSM Hamiltonian as the
$^4$He spectrum.  We use this calculated threshold in the inversion so that
the threshold, discrete spectrum, and inversion basis are defined consistently.
In some LIT applications, the calculated cross section is instead shifted to
the empirical photo-disintegration threshold when comparing with data, e.g.,
Ref.~\cite{QuaglioniNavratil2007}.  For example, for the
$N_{\max}=15$, $\hbar\Omega=15$ MeV calculation used below, the NCSM
energies give $\omega_{\rm th}=19.93$ MeV, close to the empirical
$p+{}^3{\rm H}$ separation threshold $S_p=19.81$ MeV~\cite{AME2020}.
The parameter $\beta$ sets the basis scale, and $N_b$ is the number of basis functions.
Substituting Eq.~\eqref{eq:response_expansion} into the first line of
Eq.~\eqref{eq:lit_discrete}, and interchanging the finite basis expansion
with the energy integral, leads to the linear form
\begin{equation}
L_{\rm fit}(\sigma_R,\sigma_I)
=
\sum_{n=1}^{N_b}
c_n I_n(\sigma_R,\sigma_I),
\label{eq:Lfit}
\end{equation}
where
\begin{equation}
I_n(\sigma_R,\sigma_I)
=
\int_{\omega_{\rm th}}^\infty
dE_\gamma\,
\frac{\chi_n(E_\gamma-\omega_{\rm th})}
{(E_\gamma-\sigma_R)^2+\sigma_I^2}.
\label{eq:basis_lit}
\end{equation}
The coefficients $c_n$ are determined by fitting $L_{\rm fit}$ to the finite-spectrum LIT on the sampled $\sigma_R$ grid. In the present calculations, we use a linear grid from 0 to 200 MeV with 400 points; a denser 800-point grid yields no visible change in the reconstructed cross section on the scale of the figures. The coefficients are obtained by minimizing
\begin{equation}
\sum_i
\left|
L(\sigma_R^i,\sigma_I)
-
\sum_{n=1}^{N_b}
c_n I_n(\sigma_R^i,\sigma_I)
\right|^2.
\label{eq:least_squares}
\end{equation}
Since the physical response is a sum of transition strengths, it must be
non-negative.  A non-negativity constraint is therefore imposed on the
reconstructed response over the reconstruction interval, in line with standard
LIT inversion and regularization practice
\cite{Andreasi2005,BarneaInversion2010}.
We assess the quality and stability of the inversion with three diagnostics:
the relative LIT fitting error
$\epsilon_{\rm LIT}=\|L_{\rm fit}-L\|_2/\|L\|_2$, which is minimized in
Eq.~\eqref{eq:least_squares}; the coefficient-vector norm
$\|\mathbf c\|_2=(\sum_{n=1}^{N_b}c_n^2)^{1/2}$, where the subscript 2
denotes the Euclidean 2-norm; and the peak position of the reconstructed
cross section.  The latter two quantities are not assigned fixed numerical
cutoffs, but are used as stability indicators: a reliable inversion should
avoid a rapidly growing coefficient norm and should keep the peak position
stable under variations of the inversion parameters, as illustrated below.

The photoabsorption cross section is obtained from the reconstructed response as
\begin{equation}
\sigma_\gamma(E_\gamma)
=
\frac{16\pi^3}{9}\,
\alpha_{\rm em}\,
E_\gamma R(E_\gamma),
\label{eq:sigma_gamma}
\end{equation}
where $\alpha_{\rm em}$ is the fine-structure constant.
As an internal check, we compare the $E1$ polarizability and bremsstrahlung
sum rule (BSR) obtained directly from the discrete spectrum with those from
the reconstructed cross section.
For an upper energy limit $\bar{\omega}$, the direct forms evaluated from the
discrete spectrum are
\begin{align}
\alpha_E^{\rm disc}(\bar{\omega})
&=
\frac{8\pi}{9}\,
\alpha_{\rm em}
\sum_{\omega_k\leq \bar{\omega}}
\frac{B_k(E1)}{\omega_k},
\nonumber\\
\Sigma_{\rm BSR}^{\rm disc}(\bar{\omega})
&=
\frac{16\pi^3}{9}\,
\alpha_{\rm em}
\sum_{\omega_k\leq \bar{\omega}} B_k(E1),
\label{eq:sumrules_disc}
\end{align}
where $\alpha_E$ and $\Sigma_{\rm BSR}$ denote the $E1$ polarizability and the
BSR, respectively.
The corresponding quantities obtained from the LIT-inverted photoabsorption
cross section are
\begin{align}
\alpha_E^{\rm inv}(\bar{\omega})
&=
\frac{1}{2\pi^2}
\int_{\omega_{\rm th}}^{\bar{\omega}}
dE_\gamma\,
\frac{\sigma_\gamma(E_\gamma)}{E_\gamma^2},
\nonumber\\
\Sigma_{\rm BSR}^{\rm inv}(\bar{\omega})
&=
\int_{\omega_{\rm th}}^{\bar{\omega}}
dE_\gamma\,
\frac{\sigma_\gamma(E_\gamma)}{E_\gamma}.
\label{eq:sumrules_inv}
\end{align}

\section{Results and discussion}
\label{sec:results}

We begin by establishing the convergence of the finite-spectrum LIT and the
reconstructed photoabsorption cross section with respect to the model
space.
We then examine the sensitivity of the cross section to the excitation-energy cutoff
$\omega_{\rm cut}$ and to the inversion parameters that enter the LIT
reconstruction.
The calculation is further tested through sum-rule consistency checks for the
$E1$ polarizability and BSR, and the resulting photoabsorption cross section
is finally compared with available photonuclear data and a previous
calculation with chiral interactions.

\begin{figure}[t!]
  \centering
  \includegraphics[width=1.0\linewidth]{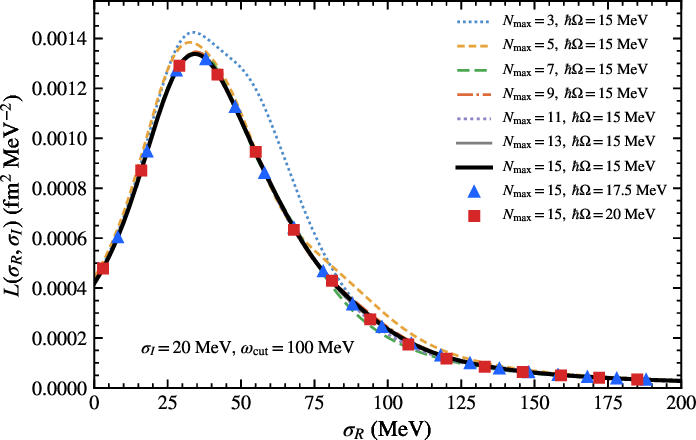}
  \caption{
(Color online) Finite-spectrum LIT $L(\sigma_R,\sigma_I)$, Eq.~\eqref{eq:lit_discrete}, of the $^{4}$He $E1$ response as a function of $\sigma_R$, calculated
with the NCSM using the Daejeon16 $NN$ interaction for $\hbar\Omega=15$ MeV
at $N_{\max}=3$--15 (lines) and for $\hbar\Omega=17.5$ and 20 MeV at
$N_{\max}=15$ (symbols), with $\sigma_I=20$ MeV and
$\omega_{\rm cut}=100$ MeV.
}
  \label{fig:lit_convergence}
\end{figure}
Before reconstructing the photoabsorption cross section, we first examine the
convergence of the finite-spectrum LIT with respect to the HO basis.
In Fig.~\ref{fig:lit_convergence}, we present the LIT $L(\sigma_R,\sigma_I)$ in Eq.~\eqref{eq:lit_discrete} for the $^{4}$He $E1$
response at $\hbar\Omega=15$ MeV with $N_{\max}=3$--15, using
$\sigma_I=20$ MeV and $\omega_{\rm cut}=100$ MeV.
We observe visible differences among the results in small model spaces.
As $N_{\max}$ increases, the curves move systematically closer to each other,
showing the convergence of the transformed strength.
The results for the largest model spaces are nearly indistinguishable on the
scale of the figure.
We also show the calculations for $\hbar\Omega=17.5$ and 20 MeV at
$N_{\max}=15$.
These results are in good agreement with the $\hbar\Omega=15$ MeV result
at the same $N_{\max}$, indicating that the $\hbar\Omega$ dependence
is weak.
This demonstrates that the finite-spectrum LIT is well converged with respect
to the HO basis before the inversion is performed.
We also checked the convergence of the LIT with respect to the
excitation-energy cutoff.  Taking the $N_{\max}=15$, $\hbar\Omega=15$ MeV
spectrum as an example, we find no visible difference between the LITs obtained
with $\omega_{\rm cut}=95$ and 100 MeV.

\begin{figure}[t!]
  \centering
  \includegraphics[width=1.0\linewidth]{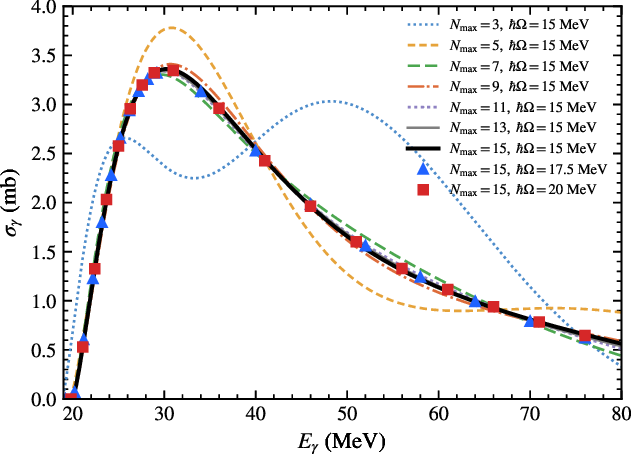}
  \caption{
  (Color online) Photoabsorption cross section of $^{4}$He obtained from the
inversion of the finite-spectrum LIT calculated with the NCSM using the
Daejeon16 $NN$ interaction for $\hOm=15$ MeV at $N_{\max}=3$--15 (lines)
and for $\hOm=17.5$ and 20 MeV at $N_{\max}=15$ (symbols), with
$\beta=4$ MeV, $\sigma_I=20$ MeV, $N_b=6$, and
$\omega_{\rm cut}=100$ MeV.
  }
  \label{fig:nmax_hw}
\end{figure}

We next examine whether the convergence observed at the LIT level is preserved
after the inversion.
In Fig.~\ref{fig:nmax_hw}, we present the reconstructed photoabsorption cross
section of $^{4}$He for $\hOm=15$ MeV with $N_{\max}=3$--15, using
$\beta=4$ MeV, $\sigma_I=20$ MeV, $N_b=6$, and $\omega_{\rm cut}=100$ MeV.
We observe sizable differences among the results in small model spaces.
With $N_{\max}$ increasing, the reconstructed cross sections exhibit a systematic convergence trend.
The results for the largest model spaces show only a weak
$N_{\max}$ dependence over the displayed energy range.
We also show the calculations for $\hOm=17.5$ and 20 MeV at $N_{\max}=15$.
These results are nearly indistinguishable from the $\hOm=15$ MeV curve on the
scale of the figure, indicating weak $\hOm$ dependence in the largest
model space.
The reconstructed cross section is therefore well converged with respect to the HO
basis.

\label{subsec:cutoff}
\begin{figure}[t!]
  \centering
  \includegraphics[width=1.0\linewidth]{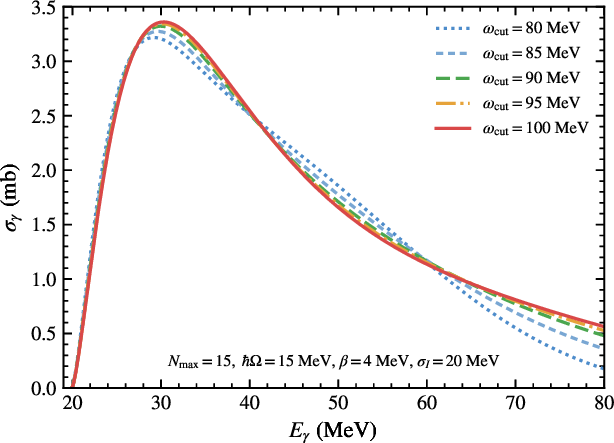}
  \caption{
  Dependence of the inverted cross section on the maximum retained excitation energy $\omega_{\rm cut}$.
  The calculation uses $N_{\max}=15$, $\hOm=15$ MeV, $\beta=4$ MeV, $\sigma_I=20$ MeV, and $N_b=6$.
  }
  \label{fig:energy_cutoff}
\end{figure}
We next examine the dependence of the cross section on the excitation-energy cutoff
$\omega_{\rm cut}$ used to truncate the finite $1^-$ spectrum.
Although the main giant-dipole strength lies well below the upper cutoff, the
Lorentz kernel has a finite width and therefore gives non-negligible weight
to higher-lying discrete states.
The dependence on $\omega_{\rm cut}$ must therefore be checked explicitly.
In Fig.~\ref{fig:energy_cutoff}, we show the photoabsorption cross sections
obtained from the spectrum at $N_{\max}=15$ with $\hbar\Omega=15$ MeV by varying
$\omega_{\rm cut}$ from 80 to 100 MeV.
The results obtained with small $\omega_{\rm cut}$ show visible deviations, indicating that the omitted
high-lying strength still has a non-negligible effect on the reconstruction.
With $\omega_{\rm cut}$ increasing, the cutoff-induced variation is strongly reduced.
In particular, the results of $\omega_{\rm cut}=95$ and 100 MeV are nearly
indistinguishable over the main response region and remain close in the
displayed high-energy region.
This cutoff plateau indicates that the retained finite spectrum up to
$\omega_{\rm cut}=100$ MeV provides a sufficiently complete input for the
present LIT inversion. We therefore use $\omega_{\rm cut}=100$ MeV in the following calculations.

\begin{figure}[t!]
  \centering
  \includegraphics[width=0.98\linewidth]{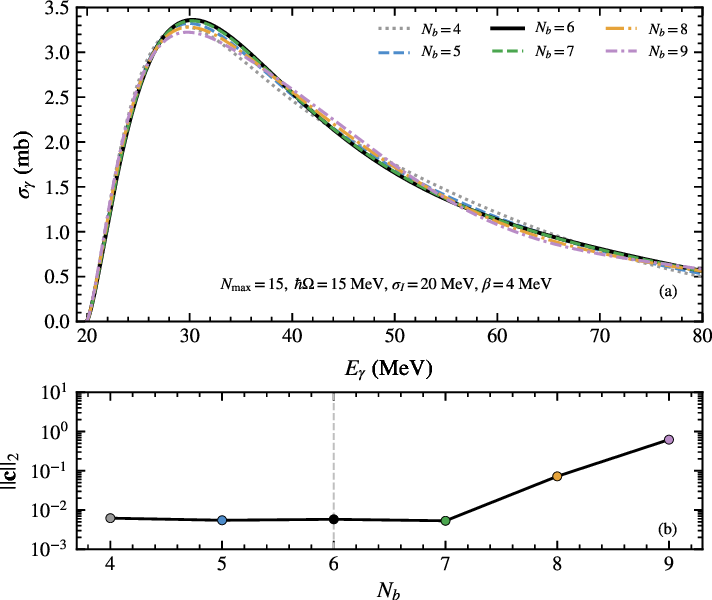}
\caption{
(Color online) Inversion stability with respect to the number of basis
functions $N_b$ for the $^{4}$He photoabsorption cross section at
$N_{\max}=15$ and $\hbar\Omega=15$ MeV.
Panel (a): the reconstructed cross sections for $N_b=4$--9;
Panel (b): the Euclidean norm of the expansion coefficients,
$\|\mathbf c\|_2$, for the same inversions.
The calculations are obtained with $\beta=4$ MeV and $\sigma_I=20$ MeV.
}
  \label{fig:Nb_stability}
\end{figure}
We next assess the stability of the LIT inversion with respect to the number
of basis functions $N_b$.
In Fig.~\ref{fig:Nb_stability}(a), we present the reconstructed
photoabsorption cross sections obtained with $N_b=4$--9 for the
$N_{\max}=15$, $\hbar\Omega=15$ MeV spectrum, using $\beta=4$ MeV and
$\sigma_I=20$ MeV.
The cross sections obtained with $N_b=5$--7 are nearly indistinguishable on
the scale of the figure, defining a stable range for the reconstruction.
By contrast, the $N_b=4$ result remains close to these curves in the resonance
region but exhibits visible oscillatory behavior above the resonance peak,
indicating that four basis functions are insufficient, whereas the
$N_b=8$ and 9 curves show increasing deviations from the stable
$N_b=5$--7 results.

In Fig.~\ref{fig:Nb_stability}(b), we show the corresponding coefficient norm
$\|\mathbf c\|_2$.
For $N_b\geq8$, we observe that $\|\mathbf c\|_2$ increases rapidly, even though the
reconstructed cross section changes only modestly relative to the stable $N_b=5$--7 results.
This indicates that the higher-$N_b$ solutions acquire increasingly large
coefficient cancellations rather than adding stable information to the
reconstruction.
We therefore use $N_b=6$ in the following calculations, because it lies within
the stable $N_b=5$--7 range and remains safely below the large-coefficient-norm
regime observed at higher $N_b$.
The remaining inversion stability is examined below through variations of
$\beta$ and $\sigma_I$.

\begin{figure*}[t!]
  \centering
  \includegraphics[width=0.98\textwidth]{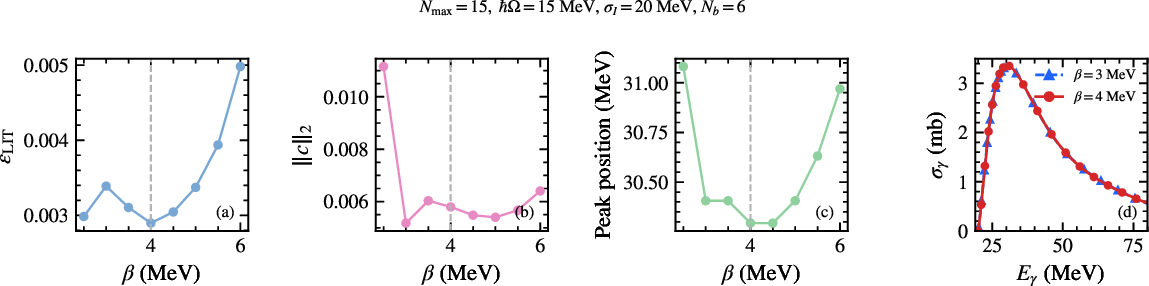}
\caption{
(Color online) Inversion stability with respect to the basis scale $\beta$
for the $^{4}$He photoabsorption cross section at $N_{\max}=15$ and
$\hbar\Omega=15$ MeV, using $N_b=6$ and $\sigma_I=20$ MeV.
Panels (a)--(c) show the relative LIT fitting error $\epsilon_{\rm LIT}$,
the coefficient norm $\|\mathbf c\|_2$, and the peak position of the
reconstructed cross section as functions of $\beta$, respectively; the
vertical dashed line marks the value $\beta=4$ MeV used in the final
reconstruction.
Panel (d) compares the reconstructed cross sections for $\beta=3$ and
4 MeV.
}
  \label{fig:e1_stability}
\end{figure*}
With $N_b=6$ and $\sigma_I=20$ MeV fixed, we examine the stability of the
LIT inversion with respect to the basis scale $\beta$ for the
$N_{\max}=15$, $\hbar\Omega=15$ MeV spectrum.
The parameter $\beta$ controls the exponential falloff of the inversion basis
functions in Eq.~\eqref{eq:basis_function}.
In Fig.~\ref{fig:e1_stability}(a)--(c), we present the relative LIT fitting
error $\epsilon_{\rm LIT}$, the coefficient norm $\|\mathbf c\|_2$, and the
peak position of the reconstructed cross section as functions of $\beta$.
A reliable choice of $\beta$ should give a small LIT fitting error, avoid a
large coefficient norm, and keep the extracted peak position stable.
The results show a stable region for $\beta=3$--5 MeV according to these three
diagnostics.
We therefore adopted $\beta=4$ MeV in this work, which lies near the center of this stable region.
This choice is further supported by Fig.~\ref{fig:e1_stability}(d), where the
reconstructed cross sections obtained with $\beta=3$ and 4 MeV are nearly
indistinguishable on the scale of the figure.

\begin{figure}[t!]
  \centering
  \includegraphics[width=0.98\linewidth]{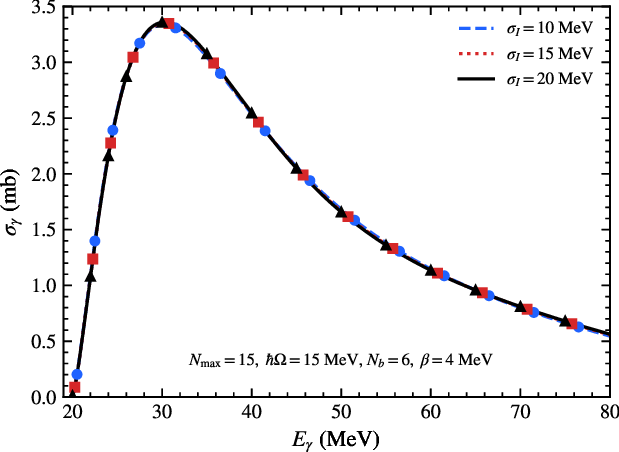}
  \caption{
  (Color online) Photoabsorption cross section of $^{4}$He reconstructed with
different Lorentz widths $\sigma_I=10$, 15, and 20 MeV for
$N_{\max}=15$ and $\hbar\Omega=15$ MeV, using $N_b=6$ and
$\beta=4$ MeV.
  }
  \label{fig:sigmaI_stability}
\end{figure}
We finally test the stability of the reconstruction against the Lorentz width
$\sigma_I$.
The parameter $\sigma_I$ controls the energy resolution of the LIT: reducing
$\sigma_I$ makes the transform more sensitive to the finite discrete spectrum,
whereas increasing it enhances the smoothing of the response.
In Fig.~\ref{fig:sigmaI_stability}, we present the reconstructed
photoabsorption cross sections obtained with $\sigma_I=10$, 15, and 20 MeV
for the $N_{\max}=15$, $\hOm=15$ MeV spectrum, with $N_b=6$ and
$\beta=4$ MeV fixed.
The three results are in close agreement over the displayed energy range,
indicating that the extracted cross section is not controlled by the particular
choice of Lorentz width within this interval.
We therefore use $\sigma_I=20$ MeV in the final reconstruction.

\begin{table}[htpb]
  \caption{
  $E1$ sum rules $\alpha_E$ and $\Sigma_{\rm BSR}$ for $^{4}$He at
$\hOm=15$ MeV, obtained directly from the discrete NCSM spectrum
(denoted by the superscript ``disc'') and from the LIT-inverted
photoabsorption cross section (denoted by the superscript ``inv'').
$\Delta$ represents the relative difference in percent.
  }
  \label{tab:alpha_consistency}
  \centering
  \scriptsize
  \setlength{\tabcolsep}{3pt}
  \resizebox{\columnwidth}{!}{%
  \begin{tabular}{@{}ccccccc@{}}
    \toprule
    $N_{\max}$ &
    \multicolumn{3}{c}{$E1$ polarizability} &
    \multicolumn{3}{c}{BSR} \\
    \cmidrule(lr){2-4}\cmidrule(l){5-7}
    &
    \begin{tabular}{@{}c@{}}$\alpha_E^{\rm disc}$\\(fm$^3$)\end{tabular} &
    \begin{tabular}{@{}c@{}}$\alpha_E^{\rm inv}$\\(fm$^3$)\end{tabular} &
    \begin{tabular}{@{}c@{}}$\Delta(\alpha_E)$\\(\%)\end{tabular} &
    \begin{tabular}{@{}c@{}}$\Sigma_{\rm BSR}^{\rm disc}$\\(mb)\end{tabular} &
    \begin{tabular}{@{}c@{}}$\Sigma_{\rm BSR}^{\rm inv}$\\(mb)\end{tabular} &
    \begin{tabular}{@{}c@{}}$\Delta(\Sigma_{\rm BSR})$\\(\%)\end{tabular} \\
    \midrule
     3 & 0.08255 & 0.08571 & 3.83 & 3.072 & 3.110 & 1.21 \\
     5 & 0.08239 & 0.08242 & 0.04 & 2.951 & 2.916 & $-1.17$ \\
     7 & 0.07807 & 0.07855 & 0.61 & 2.752 & 2.760 & 0.29 \\
     9 & 0.07831 & 0.07811 & $-0.25$ & 2.796 & 2.776 & $-0.73$ \\
    11 & 0.07811 & 0.07808 & $-0.04$ & 2.761 & 2.758 & $-0.08$ \\
    13 & 0.07824 & 0.07808 & $-0.20$ & 2.781 & 2.765 & $-0.59$ \\
    15 & 0.07832 & 0.07810 & $-0.29$ & 2.785 & 2.770 & $-0.55$ \\
    \bottomrule
  \end{tabular}%
  }
\end{table}
We use the $E1$ polarizability and BSR as internal checks of the finite-spectrum LIT inversion. For each $N_{\max}$, $\alpha_E^{\rm disc}$ and $\Sigma_{\rm BSR}^{\rm disc}$ are evaluated directly from the discrete NCSM spectrum using Eq.~\eqref{eq:sumrules_disc}, whereas $\alpha_E^{\rm inv}$ and $\Sigma_{\rm BSR}^{\rm inv}$ are obtained by integrating the reconstructed photoabsorption cross section according to Eq.~\eqref{eq:sumrules_inv}. Both the direct sums and the inverted integrals are evaluated with the same upper limit, $\bar{\omega}=100$ MeV. The matched upper limit makes this comparison an internal test of the inversion for the retained finite spectrum, rather than an estimate of the complete infinite-energy sum rules. Their agreement therefore tests whether the inversion preserves the integrated $E1$ strength within the calculated energy window. The polarizability and BSR have different sensitivity to strength beyond this window: in terms of the photoabsorption cross section, $\alpha_E$ is weighted by $E_\gamma^{-2}$, whereas the BSR is weighted only by $E_\gamma^{-1}$. Consequently, omitted high-energy strength is relatively more important for the BSR. Indeed, for $^4$He with the Daejeon16 interaction, the direct NCSM calculation of Ref.~\cite{Yin2024} estimated the omitted contribution to $\alpha_E$ from transitions above 100 MeV to be only of order $10^{-4}$ fm$^3$. Thus, while the matched-window comparison does not establish the complete BSR, the 100 MeV cutoff is sufficient for $\alpha_E$ within the corresponding excitation-energy-truncation uncertainty. Table~\ref{tab:alpha_consistency} summarizes the comparison at $\hOm=15$ MeV. The relative difference is defined as $\Delta(X)=(X^{\rm inv}-X^{\rm disc})/X^{\rm disc}$ for $X=\alpha_E$ or $\Sigma_{\rm BSR}$. The largest difference occurs at $N_{\max}=3$, where the finite spectrum gives the coarsest representation of the $E1$ strength. For $N_{\max}\geq7$, the direct and inverted values agree at the subpercent level for both $\alpha_E$ and $\Sigma_{\rm BSR}$. This agreement provides a direct validation of the reconstructed photoabsorption cross section at the sum-rule level within the calculated energy window.

As an additional closure check, we also evaluated the BSR from the ground-state expectation-value relation of Ref.~\cite{GazitSumRules2006},
\begin{equation}
\Sigma_{\rm BSR}
=
\frac{4\pi^2\alpha_{\rm em}}{3}
\left[
Z^2\langle r_p^2\rangle
-
\frac{Z(Z-1)}{2}\langle r_{pp}^2\rangle
\right].
\label{eq:bsr_density}
\end{equation}
Here $Z$ is the proton number, $\langle r_p^2\rangle$ is the mean-square point-proton radius with respect to the center of mass, and $\langle r_{pp}^2\rangle$ is the mean-square proton-proton separation. These ground-state expectation values are evaluated with the same NCSM wave function. For the Daejeon16 $NN$ interaction, this independent ground-state evaluation is well converged in the NCSM and gives $\Sigma_{\rm BSR}\simeq 2.86$ mb. This full-closure value is slightly larger than the finite-spectrum running sums in Table~\ref{tab:alpha_consistency}. The difference is expected because the table uses the same upper limit $\bar{\omega}=100$ MeV as the LIT inversion, whereas Eq.~\eqref{eq:bsr_density} represents the full closure value.

\begin{figure}[t!]
  \centering
  \includegraphics[width=0.98\linewidth]{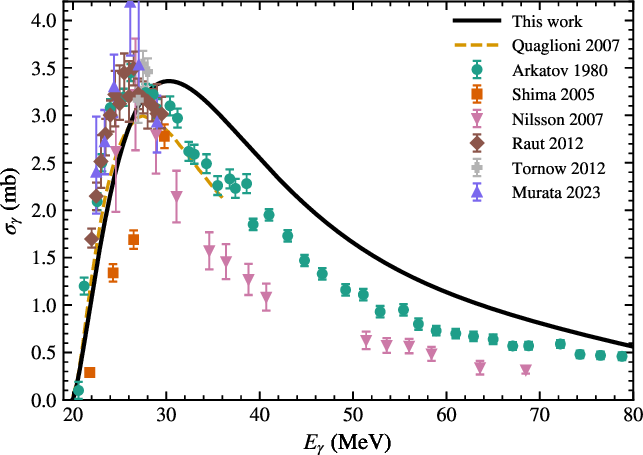}
  \caption{
  (Color online) Photoabsorption cross section of $^{4}$He compared with the
calculation of Quaglioni \textit{et al.}~\cite{QuaglioniNavratil2007} and selected
photonuclear data~\cite{ArkatovData,Shima2005,Nilsson2007,Raut2012,Tornow2012,Murata2023}.
  }
  \label{fig:experiment}
\end{figure}

In Fig.~\ref{fig:experiment}, we compare the reconstructed
photoabsorption cross section obtained at $N_{\max}=15$ and
$\hbar\Omega=15$ MeV with available photonuclear data and with the
previous calculation by Quaglioni \textit{et al.}~\cite{QuaglioniNavratil2007}.
The calculation of Quaglioni \textit{et al.} employed the NCSM with the Entem--Machleidt N$^3$LO $NN$ interaction~\cite{EntemMachleidt2003}
supplemented by the chiral N$^2$LO three-nucleon force, with the low-energy constants
of the three-nucleon interaction chosen as in Ref.~\cite{Navratil2007PRL}.
This Hamiltonian is different from the Daejeon16 interaction used in the
present work. Although Daejeon16 is also based on the Entem--Machleidt
N$^3$LO interaction, it is obtained through SRG evolution followed by
phase-equivalent transformations whose parameters were constrained by selected
light-nucleus observables~\cite{Daejeon16}.  Accordingly, contributions
associated with three-nucleon dynamics are represented differently in the two
calculations: explicitly in the Hamiltonian of Quaglioni \textit{et al.}, and
through the fitted off-shell modifications to the two-body interaction in Daejeon16.
The comparison in Fig.~\ref{fig:experiment} therefore reflects both the
different Hamiltonians and the different implementations of the LIT calculation.
The experimental data shown in the figure include both total cross sections
and exclusive two-body breakup measurements.
The Arkatov~\cite{ArkatovData} and Shima~\cite{Shima2005} data are shown as
total photoabsorption measurements.
The Raut~\cite{Raut2012} data correspond to the
$^{4}{\rm He}(\gamma,p)^{3}{\rm H}$ channel, whereas the
Nilsson~\cite{Nilsson2007} and Tornow~\cite{Tornow2012} data correspond to
the $^{4}{\rm He}(\gamma,n)^{3}{\rm He}$ channel.
For these single-channel two-body data, we multiply the measured cross
sections by two only to estimate the summed mirror two-body contribution.
The Murata~\cite{Murata2023} data correspond to the measured sum of the
$^{4}{\rm He}(\gamma,p)^{3}{\rm H}$ and
$^{4}{\rm He}(\gamma,n)^{3}{\rm He}$ channels.
As emphasized in Ref.~\cite{QuaglioniNavratil2007}, this mirror-doubling
estimate is a safe proxy for the total cross section only below the
three-body breakup threshold; above that energy it does not contain the
$^{4}{\rm He}(\gamma,np)d$ and four-body breakup channels and should be
regarded as a partial contribution rather than a total cross section.

We observe in Fig.~\ref{fig:experiment} that the present calculation captures the main features of the $^{4}$He
giant-dipole response.
It approximately follows the low-energy rise and the main resonance region of the Quaglioni
calculation, while a substantial difference appears on the
high-energy side of the resonance.
In this region, the present Daejeon16 result shows a more pronounced shoulder,
indicating a redistribution of $E1$ strength toward higher excitation
energies.
The convergence, cutoff, inversion-stability, and sum-rule checks discussed
above show that this feature is stable within the present finite-spectrum LIT
calculation.
The result therefore highlights the sensitivity of the continuum $E1$ response
to the underlying Hamiltonian.

\section{Summary}
\label{sec:summary}

We have developed a finite-spectrum implementation of the LIT within the
\textit{ab initio} NCSM and applied it to the total photoabsorption cross
section of $^{4}$He.
In this formulation, the LIT is constructed directly from the excitation
energies of explicitly calculated $J^\pi=1^-$ states and their
$E1$ transition strengths.
The smooth response is then obtained by inverting the finite-spectrum transform
and is converted to the photoabsorption cross section.

Using the Daejeon16 $NN$ interaction, we examined the convergence of both the LIT
and the reconstructed cross section with respect to the HO basis.
The results show systematic convergence with increasing $N_{\max}$, and the
$\hOm$ dependence is weak in the largest model space considered.
The dependence on the excitation-energy cutoff was also quantified, and
$\omega_{\rm cut}=100$ MeV was found to provide a sufficiently complete
finite-spectrum input for the present calculation.
Independent scans of the inversion parameters, including $N_b$, $\beta$, and
$\sigma_I$, demonstrate that the extracted cross section is stable within the
chosen reconstruction setup.

The reliability of the inversion was further tested through the $E1$
polarizability and BSR.
For $N_{\max}\geq7$, the values obtained by integrating the reconstructed
photoabsorption cross section agree at the subpercent level with the direct
finite-spectrum sums.
This agreement provides a direct validation of the reconstructed cross section
at the sum-rule level.

The final cross section describes the characteristic broad giant-dipole
structure of $^{4}$He and follows the overall trend of the available
photonuclear data.
When we compare our cross section with a previous calculation based on chiral interactions that directly include three-nucleon interactions, we find both give a sharp low-energy rise with $~2$ MeV difference in peak locations and a significantly
different high-energy strength distribution, which reveal the sensitivity
of the continuum $E1$ response to the underlying Hamiltonian.
Overall, the present work provides a controlled route from explicitly computed
NCSM spectra and transition strengths to photonuclear observables.

\begin{acknowledgments}
We acknowledge helpful discussions with Pieter Maris and Mark A. Caprio.
This work is partially supported by Natural Science Foundation of Henan under Grant No.~252300421486, by the State Key Laboratory of Heavy Ion Science and Technology, Institute of Modern Physics, Chinese Academy of Sciences (Grant No.~HIST2026CO09), and National Key R\&D Program of China (2023YFA1606701).
A portion of the computational resources are provided by the National Energy Research Scientific Computing Center (NERSC), a U.S. Department of Energy Office of Science User Facility located at Lawrence Berkeley National Laboratory, operated under Contract No. DE-AC02-05CH11231 using NERSC awards NP-ERCAP0020944, NP-ERCAP0023866 and NP-ERCAP0028672.
This work is partially supported
by new faculty startup funding by the Institute of Modern Physics, Chinese Academy of Sciences, by Key Research Program of Frontier Sciences, Chinese Academy
of Sciences, Grant No. ZDB-SLY-7020, by the Natural
Science Foundation of Gansu Province, China, Grant
No.~20JR10RA067, by the Foundation for Key Talents of Gansu Province, by the Central Funds Guiding the Local Science and Technology Development of Gansu Province, Grant No. 22ZY1QA006, by international partnership program of the Chinese Academy
of Sciences, Grant No. 016GJHZ2022103FN, by National Natural Science Foundation of China, Grant No. 12375143, by National Key R\&D Program of China,
Grant No. 2023YFA1606903 and by the Strategic Priority Research Program of the Chinese Academy of Sciences, Grant No. XDB34000000.
P.~Yin and B.~Zhou are supported by Shanghai Research Center for Theoretical Nuclear Physics, NSFC and Fudan University, Shanghai 200438, China and the National Natural Science Foundation of China under Grant No.12147101.
A.~M.~Shirokov is thankful to the Chinese Academy of Sciences President's International Fellowship Initiative Program (Grant No.~2023VMA0013) which supported his visits to Lanzhou where a part of this work was performed and acknowledges the hospitality of Chinese colleagues during these visits.
A portion of the computational resources were also provided by Gansu Computing Center and Gansu Advanced Computing Center. This material is based upon work supported by the U.S.~Department of Energy, Office of Science, under Award No.~DE-FG02-95ER40934.
\end{acknowledgments}


\begin{thebibliography}{99}
\bibitem{EfrosLITReview}
V.~D. Efros, W.~Leidemann, G.~Orlandini, and N.~Barnea,
``The Lorentz integral transform (LIT) method and its applications to perturbation-induced reactions,''
J. Phys. G \textbf{34}, R459 (2007).
\bibitem{LeidemannOrlandini2013}
W.~Leidemann and G.~Orlandini,
``Modern ab initio approaches and applications in few-nucleon physics with $A\geq4$,''
Prog. Part. Nucl. Phys. \textbf{68}, 158 (2013).
\bibitem{BaccaPastore2014}
S.~Bacca and S.~Pastore,
``Electromagnetic reactions on light nuclei,''
J. Phys. G \textbf{41}, 123002 (2014).
\bibitem{BermanFultz1975}
B.~L. Berman and S.~C. Fultz,
``Measurements of the giant dipole resonance with monoenergetic photons,''
Rev. Mod. Phys. \textbf{47}, 713 (1975).
\bibitem{HarakehWoude2001}
M.~N. Harakeh and A.~van der Woude,
\textit{Giant Resonances: Fundamental High-Frequency Modes of Nuclear Excitation}
(Oxford University Press, Oxford, 2001).
\bibitem{GazitSumRules2006}
D.~Gazit, N.~Barnea, S.~Bacca, W.~Leidemann, and G.~Orlandini,
``Photonuclear sum rules and the tetrahedral configuration of $^{4}$He,''
Phys. Rev. C \textbf{74}, 061001(R) (2006).
\bibitem{Stetcu2007}
I.~Stetcu, S.~Quaglioni, S.~Bacca, B.~R. Barrett, C.~W. Johnson, P.~Navr\'atil, N.~Barnea, W.~Leidemann, and G.~Orlandini,
``Benchmark calculation of inclusive electromagnetic responses in the four-body nuclear system,''
Nucl. Phys. A \textbf{785}, 307 (2007).
\bibitem{QuaglioniNavratil2007}
S.~Quaglioni and P.~Navr\'atil,
``The $^{4}$He total photo-absorption cross section with two- plus three-nucleon interactions from chiral effective field theory,''
Phys. Lett. B \textbf{652}, 370 (2007).

\bibitem{KegelMonopole2023}
S.~Kegel \textit{et al.},
``Measurement of the $\alpha$-particle monopole transition form factor challenges theory: A low-energy puzzle for nuclear forces?''
Phys. Rev. Lett. \textbf{130}, 152502 (2023).
\bibitem{BaccaMonopole2013}
S.~Bacca, N.~Barnea, W.~Leidemann, and G.~Orlandini,
``The isoscalar monopole resonance of the alpha particle: A prism to nuclear Hamiltonians,''
Phys. Rev. Lett. \textbf{110}, 042503 (2013).
\bibitem{Michel2023}
N.~Michel, W.~Nazarewicz, and M.~P\l{}oszajczak,
``Description of the proton-decaying $0_2^+$ resonance of the $\alpha$ particle,''
Phys. Rev. Lett. \textbf{131}, 242502 (2023);
Erratum, Phys. Rev. Lett. \textbf{133}, 239901 (2024).
\bibitem{Meissner2024}
U.-G.~Mei\ss{}ner, S.~Shen, S.~Elhatisari, and D.~Lee,
``Ab initio calculation of the alpha-particle monopole transition form factor,''
Phys. Rev. Lett. \textbf{132}, 062501 (2024).
\bibitem{Viviani2024}
M.~Viviani, A.~Kievsky, L.~E. Marcucci, and L.~Girlanda,
``Study of the alpha-particle monopole transition form factor,''
Few-Body Syst. \textbf{65}, 74 (2024).
\bibitem{YinMonopole2025}
P.~Yin, A.~M. Shirokov, H.~Li, B.~Zhou, X.~Zhao, S.~Bacca, and J.~P. Vary,
``$\alpha$-particle monopole form factors within the ab initio no-core shell model,''
Phys. Rev. C \textbf{112}, L031303 (2025).
\bibitem{Gazit2006}
D.~Gazit, S.~Bacca, N.~Barnea, W.~Leidemann, and G.~Orlandini,
``Photoabsorption on 4He with a Realistic Nuclear Force,''
Phys. Rev. Lett. \textbf{96}, 112301 (2006).
\bibitem{ArkatovData}
Y.~M. Arkatov, P.~I. Vatset, V.~I. Voloshchuk,
V.~A. Zolenko, and I.~M. Prokhorets,
``Experimental verification of the sum rules for photodisintegration of He-4,''
Yad. Fiz. \textbf{31}, 1400 (1980)
[Sov. J. Nucl. Phys. \textbf{31}, 726 (1980)].
\bibitem{Calarco1983}
J.~R. Calarco, B.~L. Berman, and T.~W. Donnelly,
``Implications of the experimental results on the photodisintegration of $^{4}$He,''
Phys. Rev. C \textbf{27}, 1866 (1983).
\bibitem{Shima2005}
T.~Shima, S.~Naito, Y.~Nagai, T.~Baba, K.~Tamura, T.~Takahashi, T.~Kii, H.~Ohgaki, and H.~Toyokawa,
``Simultaneous measurement of the photodisintegration of $^{4}$He in the giant dipole resonance region,''
Phys. Rev. C \textbf{72}, 044004 (2005).
\bibitem{Nilsson2007}
B.~Nilsson \textit{et al.},
``A measurement of the $^{4}$He($\gamma,n$) reaction from $23<E_\gamma<70$ MeV,''
Phys. Rev. C \textbf{75}, 014007 (2007).
\bibitem{Raut2012}
R.~Raut \textit{et al.},
``Photodisintegration cross section of the reaction $^{4}$He($\gamma,p$)$^{3}$H at the giant dipole resonance peak,''
Phys. Rev. Lett. \textbf{108}, 042502 (2012).
\bibitem{Tornow2012}
W.~Tornow, J.~H. Kelley, R.~Raut, G.~Rusev, A.~P. Tonchev, M.~W. Ahmed, A.~S. Crowell, and S.~C. Stave,
``Photodisintegration cross section of the reaction $^{4}$He($\gamma,n$)$^{3}$He at the giant dipole resonance peak,''
Phys. Rev. C \textbf{85}, 061001(R) (2012).
\bibitem{Murata2023}
M.~Murata \textit{et al.},
``Photodisintegration cross section of $^{4}$He in the giant dipole resonance energy region,''
Phys. Rev. C \textbf{107}, 064317 (2023).
\bibitem{EfrosLIT}
V.~D. Efros, W.~Leidemann, and G.~Orlandini,
``Response functions from integral transforms with a Lorentz kernel,''
Phys. Lett. B \textbf{338}, 130 (1994).
\bibitem{Andreasi2005}
D.~Andreasi, W.~Leidemann, C.~Reiss, and M.~Schwamb,
``New inversion methods for the Lorentz integral transform,''
Eur. Phys. J. A \textbf{24}, 361 (2005).
\bibitem{BarneaInversion2010}
N.~Barnea, V.~D. Efros, W.~Leidemann, and G.~Orlandini,
``The Lorentz integral transform and its inversion,''
Few-Body Syst. \textbf{47}, 201 (2010).
\bibitem{Leidemann2015}
W.~Leidemann,
``Energy resolution with the Lorentz integral transform,''
Phys. Rev. C \textbf{91}, 054001 (2015).
\bibitem{MarchisioLanczos}
M.~A. Marchisio, N.~Barnea, W.~Leidemann, and G.~Orlandini,
``Efficient Method for Lorentz Integral Transforms of Reaction Cross Sections,''
Few-Body Syst. \textbf{33}, 259 (2003).
\bibitem{Bacca2002}
S.~Bacca, M.~A. Marchisio, N.~Barnea, W.~Leidemann, and G.~Orlandini,
``Microscopic calculation of six-body inelastic reactions with complete final state interaction: Photoabsorption of $^{6}$He and $^{6}$Li,''
Phys. Rev. Lett. \textbf{89}, 052502 (2002).
\bibitem{Barnea2004}
S.~Bacca, N.~Barnea, W.~Leidemann, and G.~Orlandini,
``Effect of $P$-wave interaction in $^{6}$He and $^{6}$Li photoabsorption,''
Phys. Rev. C \textbf{69}, 057001 (2004).
\bibitem{Schuster2014}
M.~D. Schuster, S.~Quaglioni, C.~W. Johnson, E.~D. Jurgenson, and P.~Navr\'atil,
``Operator evolution for ab initio electric dipole transitions of $^{4}$He,''
Phys. Rev. C \textbf{92}, 014320 (2015).
\bibitem{BaccaCC2013}
S.~Bacca, N.~Barnea, G.~Hagen, G.~Orlandini, and T.~Papenbrock,
``First principles description of the giant dipole resonance in $^{16}$O,''
Phys. Rev. Lett. \textbf{111}, 122502 (2013).
\bibitem{Miorelli2016}
M.~Miorelli, S.~Bacca, N.~Barnea, G.~Hagen, G.~R. Jansen, G.~Orlandini, and T.~Papenbrock,
``Electric dipole polarizability from first principles calculations,''
Phys. Rev. C \textbf{94}, 034317 (2016).
\bibitem{OrlandiniCC2014}
G.~Orlandini, S.~Bacca, N.~Barnea, G.~Hagen, M.~Miorelli, and T.~Papenbrock,
``Coupling the Lorentz integral transform (LIT) and the coupled cluster (CC) methods: A way towards continuum spectra of ``not-so-few-body'' systems,''
Few-Body Syst. \textbf{55}, 907 (2014).
\bibitem{Yin2024}
P.~Yin, A.~M. Shirokov, P.~Maris, P.~J. Fasano, M.~A. Caprio, H.~Li, W.~Zuo, and J.~P. Vary,
``Direct ab initio calculation of the $^{4}$He nuclear electric dipole polarizability,''
Phys. Lett. B \textbf{855}, 138857 (2024).
\bibitem{Barrett2013}
B.~R. Barrett, P.~Navr\'atil, and J.~P. Vary,
``Ab initio no core shell model,''
Prog. Part. Nucl. Phys. \textbf{69}, 131 (2013).
\bibitem{Maris2016}
P.~Maris \textit{et al.},
``Properties of $^{4}$He and $^{6}$Li with improved chiral EFT interactions,''
EPJ Web Conf. \textbf{113}, 04015 (2016).
\bibitem{Maris2021}
P.~Maris \textit{et al.} (LENPIC Collaboration),
``Light nuclei with semilocal momentum-space regularized chiral interactions up to third order,''
Phys. Rev. C \textbf{103}, 054001 (2021).
\bibitem{Maris2022}
P.~Maris \textit{et al.} (LENPIC Collaboration),
``Nuclear properties with semilocal momentum-space regularized chiral interactions beyond N$^2$LO,''
Phys. Rev. C \textbf{106}, 064002 (2022).
\bibitem{LiCPC2024}
H.~Li, H.~J. Ong, D.-L. Fang, I.~A. Mazur, I.~J. Shin, A.~M. Shirokov, J.~P. Vary, P.~Yin, X.-B. Zhao, and W.~Zuo,
``Ab initio study of $Z(N)=6$ magicity,''
Chin. Phys. C \textbf{48}, 124103 (2024).
\bibitem{LiPRC2024}
H.~Li, D.~Fang, H.~J. Ong, A.~M. Shirokov, J.~P. Vary, P.~Yin, and X.~Zhao,
``Quadrupole dynamics of carbon isotopes and $^{10}$Be,''
Phys. Rev. C \textbf{110}, 064325 (2024).
\bibitem{HuangHalo2025}
M.~Huang, T.~Frederico, P.~Yin, R.~A.~M.~Basili, P.~J.~Fasano and J.~P.~Vary,
``Halo structure of 6He from ab initio two-nucleon spatial correlations,''
Phys. Rev. C \textbf{113}, 064318 (2026).
\bibitem{Daejeon16}
A.~M. Shirokov, I.~J. Shin, Y.~Kim, M.~Sosonkina, P.~Maris, and J.~P. Vary,
``N3LO NN interaction adjusted to light nuclei in ab exitu approach,''
Phys. Lett. B \textbf{761}, 87 (2016).
\bibitem{MarisMFDn2010}
P.~Maris, M.~Sosonkina, J.~P.~Vary, E.~Ng, and C.~Yang,
``Scaling of ab-initio nuclear physics calculations on multicore computer architectures,''
Procedia Comput. Sci. \textbf{1}, 97--106 (2010).
\bibitem{AktulgaMFDn2014}
H.~M. Aktulga, C.~Yang, E.~G. Ng, P.~Maris, and J.~P. Vary,
``Improving the scalability of a symmetric iterative eigensolver for multi-core platforms,''
Concurr. Comput. Pract. Exp. \textbf{26}, 2631--2651 (2014).
\bibitem{ShaoMFDn2018}
M.~Shao, H.~M. Aktulga, C.~Yang, E.~G. Ng, P.~Maris, and J.~P. Vary,
``Accelerating nuclear configuration interaction calculations through a preconditioned block iterative eigensolver,''
Comput. Phys. Commun. \textbf{222}, 1--13 (2018).
\bibitem{MarisMFDnGPU2022}
P.~Maris, C.~Yang, D.~Oryspayev, and B.~Cook,
``Accelerating an iterative eigensolver for nuclear structure configuration interaction calculations on GPUs using OpenACC,''
J. Comput. Sci. \textbf{59}, 101554 (2022).
\bibitem{CookMFDn2022}
B.~G. Cook, P.~J. Fasano, P.~Maris, C.~Yang, and D.~Oryspayev,
``Accelerating quantum many-body configuration interaction with directives,''
Lect. Notes Comput. Sci. \textbf{13194}, 112--132 (2022).
\bibitem{FasanoMFDnTransitions2025}
P.~J. Fasano and P.~Maris,
\textsc{mfdn-transitions}, version 1.0.0,
Zenodo (2025),
doi:10.5281/zenodo.18013362.
\bibitem{GloecknerLawson}
D.~H. Gloeckner and R.~D. Lawson,
``Spurious center-of-mass motion,''
Phys. Lett. B \textbf{53}, 313 (1974).
\bibitem{Whitehead1977}
R.~R. Whitehead, A.~Watt, B.~J. Cole, and I.~Morrison,
``Computational methods for shell-model calculations,''
Adv. Nucl. Phys. \textbf{9}, 123 (1977).



\bibitem{AME2020}
M.~Wang, W.~J. Huang, F.~G. Kondev, G.~Audi, and S.~Naimi,
``The AME 2020 atomic mass evaluation (II). Tables, graphs and references,''
Chin. Phys. C \textbf{45}, 030003 (2021).
\bibitem{EntemMachleidt2003}
D.~R. Entem and R.~Machleidt,
``Accurate charge-dependent nucleon-nucleon potential at fourth order of chiral perturbation theory,''
Phys. Rev. C \textbf{68}, 041001(R) (2003).

\bibitem{Navratil2007PRL}
P.~Navr\'atil, V.~G. Gueorguiev, J.~P. Vary, W.~E. Ormand, and A.~Nogga,
``Structure of $A=10$--13 nuclei with two- plus three-nucleon interactions from chiral effective field theory,''
Phys. Rev. Lett. \textbf{99}, 042501 (2007).

\end{thebibliography}
\end{document}